\documentclass[aps,pre,twocolumn,showpacs,superscriptaddress]{revtex4}
\usepackage[latin1]{inputenc}
\usepackage[english]{babel}
\usepackage{bm,graphicx,graphics,amsmath,amssymb,epsfig}
\usepackage{txfonts}

\usepackage[colorlinks,citecolor=blue,linkcolor=Fuchsia]{hyperref}
\usepackage[usenames,dvipsnames,svgnames]{xcolor}
\usepackage{commath}

\def \be {\begin{equation}}
\def \ee {\end{equation}}
\def \beA {\begin{eqnarray}}
\def \eeA {\end{eqnarray}}

\def \der {\partial}

\def \Re  {\rm{Re}}
\def \Im  {\rm{Im}}
\def \average#1{\left\langle #1 \right\rangle}

\def \de  {\rm{d}} 

\begin{document}

\title{Information transfer in coupled Langevin equations}
\author{Simone Borlenghi} 
\affiliation{Department of Applied Physics, School of Engineering Science, KTH Royal Institute of Technology, Electrum 229, SE-16440 Kista, Sweden}
\date{\today}

\begin{abstract}
We provide a general formula, based on stochastic thermodynamics, that describes the flow of information between an arbitrary number of coupled complex-valued Langevin equations. This permits to describe the transfer of information in complex networks of oscillators out of thermal equilibrium, that can model a multitude of physical, biological and man made systems. The information flow contains an incoherent component proportional to the amplitude difference and a coherent one proportional to the phase difference between the oscillators, which depends on their synchronisation.
We illustrate the theory by simulating the dynamics of a spin-Seebeck diode, described by two coupled oscillators, that can rectify the flow of information, energy and spin. Remarkably, the system can operate in a regime where the synchronisation is broken and there is a flow of incoherent information without net transfer of energy.
\end{abstract}

\maketitle 

Every complex physical system produces and transfers information among its subparts. A precise definition of  information transfer, called transfer entropy \cite{schreiber00} was formulated in the early 2000 and has been fundamental in quantifying the statistical coherence and the mutual influence of systems evolving in time. Although very useful and successful in several research fields, including among others neuroscience \cite{vicente11}, financial time series analysis \cite{marschinski02} and complex networks of oscillators \cite{kirst16}, this quantity is essentially a black box that provides no information on the physical process that generates it. In addition, one needs to know the probability distribution associated to the dynamical process to calculate the transfer entropy, a quantity can be difficult and computationally costly to obtain from the trajectory of a dynamical system. 

In dissipative systems out of thermal equilibrium, described by master and Langevin equations, the notion of information flow has been formulated by several authors using the formalism of Stochastic Thermodynamics (ST) \cite{seifert12,horowitz14,barato14}, and plays a pivotal role in the foundations of the thermodynamics of small systems \cite{parrondo15}. The ST formulation of information flow has allowed to relate the transfer of information to the mathematical structure of the stochastic processes that generate it and to derive fluctuation theorems associated to these processes \cite{rosinberg16}. However, the research performed so far based on ST concerning coupled Langevin equations focuses mainly on specific examples with only two coupled systems \cite{allahverdyan09,horowitz14}. A more general approach for multipartite systems, developed by Horowitz \cite{horowitz15} also requires
the knowledge of the off-equilibrium probability distribution and a different route, explored by Liang and Kleeman \cite{liang13} suffers from similar limitations.

In this Letter, we use the ST formalism to derive a simple and general formula for the information flow between an arbitrary number of coupled systems described by complex-valued Langevin equations. This permits to capture in full generality the dynamics of complex networks of nonlinear oscillators, which find application in a multitude of physical, biological and technological systems. Moreover, those systems can be driven by two thermodynamical forces, notably the difference of temperature and chemical potential, and therefore exhibit a rich dynamics with transport of coupled energy and particle currents \cite{iubini12,iubini13}. Our route is grounded on previous research on ST both by the present and other authors \cite{tome10,horowitz14,horowitz15,borlenghi17,borlenghi18}. Here, a simple algebraic passage allows one to obtain the information flow from the average of a stochastic trajectory, without the need to know the underlying probability distribution explicitly.

We find in particular that the information flow splits into two components, an incoherent component that depends only on the difference between the amplitudes of the oscillators and a coherent one that depends on their phase differences and synchronisation. 
The exchange of information due to phase synchronisation has been described in various oscillators' networks \cite{bollt12,kirst16} and neural circuits \cite{terwal17}, and our model provides a theoretical ground for these observations.

By means of simple numerical simulations of two coupled equations, that are usually adopted to model spin transfer nano oscillators (STNOs) \cite{slavin09}, we show that the information flow can be rectified in a way similar to the energy and spin-wave flows in the spin-Seebeck diode \cite{borlenghi14a,borlenghi14b}. Moreover, the system can operate in a regime were there is transfer of incoherent information without synchronisation, and thus with no net transfer of energy or spin current. 

We start by considering two coupled systems, described by the following complex-valued Langevin equations

\beA
\label{eq:langevin}
\dot{\psi}_1 &=& F_1+G_{12}+\sqrt{D_1}\xi_1\nonumber\\
\dot{\psi}_2 &=& F_2+G_{21}+\sqrt{D_2}\xi_2
\eeA
where $F_{i}$ is the force acting separately on each system, $G_{ij}$, $i,j=1,2$  is the (possibly asymmetric) coupling between them and $\xi_i$ is a complex Gaussian random
variable with zero average and correlation $\average{\xi_i(t)\xi^*_j(t^\prime)}=\delta_{ij}\delta(t-t^\prime)$. From hereon the $*$ indicates complex conjugation. The diffusion constant $D_i=\alpha_iT_i$ accounts for the strength of the fluctuations and is equal to the product of the damping coefficient $\alpha_i$ and temperature $T_i$.

Both $F_i$ and $G_{ij}$ can be arbitrary functions of the $\psi$s, the only assumption we made is that they contain a term proportional to $\alpha_i\psi_i$, so that they satisfy a 
fluctuation-dissipation theorem and they can reach thermal equilibrium with their baths when the temperatures and chemical potentials are the same.
Note that the complex variables $\psi_i(t)=\sqrt{p_i}(t)e^{i\phi_i(t)}$ can be written in terms of the powers $p_i$ and phases $\phi_i$, $i=1,2$, so that each one of Eqs.(\ref{eq:langevin}) becomes two coupled equations for the evolution of phase and amplitudes \cite{iubini12,iubini13,borlenghi14b}.

When the coupling is zero, the two systems do not interact, and their joint probability density function $P_{12}\equiv P(\psi_1,\psi_2)$ factorises into the product of the two independent probabilities, $P_{12}=P_1P_2\equiv P(\psi_1)P(\psi_2)$. Each probability evolves separately according to the following Fokker-Planck (FP) equation \cite{borlenghi17,borlenghi18}:
\label{eq:fp1}
\be
\dot{P}_i=-\der_i(F_iP_i)-\der_i^*(F_i^*P_i)+2\der_i\der_i^*P_i.
\ee
Here $\der_i\equiv\frac{\der}{\de\psi_i}$ are the Wirtinger derivatives, with $\psi_i=x_i+iy_i$, $\frac{\der}{\der\psi_i}=\frac{1}{2}\left(\frac{\der}{\der x_i} - i\frac{\der}{\der y_i}\right)$ and 
$\der_i^*$ the complex conjugate.
 
On the other hand, where the two equations are coupled, the probability $P_{12}\equiv P(\psi_1,\psi_2)$ does not factorise and satisfies the following FP equation:

\beA
\label{eq:fp2}
\dot{P}_{12} &=& -\der_1[(F_1+G_{12})P_{12}]-\der_1^*[(F^*_1+G^*_{12})P_{12}]\nonumber\\
                     &-&\der_2[(F_2+G_{21})P_{12}]-\der_2^*[(F^*_2+G^*_{21})P_{21}]\nonumber\\
                     &+&2\der_1\der_1^*P_{12}+2\der_2\der_2^*P_{12}.
\eeA
At this point it is convenient to introduce the probability currents
\beA
\label{eq:probcurrents1}
\mathcal{J}_i &=& F_iP_i-D_i\der_i^*P_i\\
\mathcal{J}^{\rm{int}}_1 &=& (F_{1}+G_{12})P_{12}-D_1\der_1^*P_{12}
\eeA
respectively for the two disjoint and interacting systems. Note that the current $\mathcal{J}_2^{\rm{int}}$ is obtained from $\mathcal{J}_1^{\rm{int}}$ by simply swapping the indexes 1 and 2.

In terms of the probability currents, the two FP equations assume the form of continuity equations \cite{tome10,borlenghi17,borlenghi18}, respectively 
\beA
\label{eq:probcurrents2}
\dot{P}_i &=& -2{\Re}\left[\der_i\mathcal{J}_i\right],\nonumber\\
\dot{P}_{12} &=& -2{\Re}\left[\der_1\mathcal{J}^{\rm{int}}_1+\der_2\mathcal{J}^{\rm{int}}_2\right].
\eeA
To calculate the information flow between the two systems, we start from the definition of mutual information 

\be
\label{eq:mutualinfo1}
\mathcal{M}={\int}P_{12}{\ln}\frac{P_{12}}{P_1P_2}dx, 
\ee
where $dx=\left(\frac{i}{2}\right)^2\prod_{i=1,2}d\psi_i\wedge d\psi_i^*$ is the phase space volume element, and calculate its time derivative:

\beA
\label{eq:mutualinfo1}
\dot{\mathcal{M}} &=& \frac{d}{dt}\int P_{12}{\ln}\frac{P_{12}}{P_1P_2} dx \nonumber\\ 
                 &=&\frac{d}{dt}\int P_{12}\ln P_{12}dx-\frac{d}{dt}\int P_{12}\ln {P_1P_2}dx.
\eeA
Upon discarding boundary terms as in Refs.\cite{borlenghi17,borlenghi18}, a straightforward calculation shows that
$\dot{\mathcal{M}}=\mathcal{I}_1+\mathcal{I}_2+\mathcal{I}_3$ is the sum of the following three integrals:
$\mathcal{I}_1=\int \dot{P}_{12}\ln P_{12}dx$, $\mathcal{I}_2= -\int \dot{P}_{12}\ln {P_1P_2}dx$ and
$\mathcal{I}_3= -\int P_{12}\left(\frac{\dot{P}_1}{P_1}+\frac{\dot{P}_2}{P_2}\right)dx$.

We see here immediately that $\mathcal{I}_3$ is a constant term that does not change upon swapping the indexes
1 and 2. Thus, it does not provide any information about the asymmetric net transfer of information and can be discarded. 

We proceed now as in Refs.\cite{tome10,borlenghi17}, by substituting $(\dot{P}_{12},\dot{P}_1,\dot{P}_2)$ with the FP equation 
Eq.(\ref{eq:probcurrents2}). This gives $\mathcal{I}_1=-2{\Re}\int(\der_1\mathcal{J}_1^{\rm{int}}+\der_2\mathcal{J}_2^{\rm{int}})\ln P_{12}dx$
 and $\mathcal{I}_2= -2{\Re}\int(\der_1\mathcal{J}_1^{\rm{int}}+\der_2\mathcal{J}_2^{\rm{int}})\ln {P_1P_2}dx$. At this point we integrate by
part and we use the relation, taken from Eqs.(\ref{eq:probcurrents1}) and (\ref{eq:probcurrents2}), 

\beA
\label{eq:usefulrelation}
\der_i \ln P_i\equiv\frac{\der_iP_i}{P_i} &=& \frac{1}{D_i}\left(F_i^*-\frac{\mathcal{J}_i^*}{P_i}\right) \\
\der_i \ln P_{12}\equiv\frac{\der_iP_{12}}{P_{12}} &=& \frac{1}{D_i}\left(F_i^*+G_{ij}^*-\frac{\mathcal{J}^{\rm{int}*}_i}{P_{12}}\right)
\eeA
for $i=1,2$. Inserting this into $\mathcal{I}_1$ and $\mathcal{I}_2$ gives:

\beA
\label{eq:mutualinfo2}
\dot{\mathcal{M}} &=& 2{\Re}\int\left(\mathcal{J}^{\rm{int}}_1G^*_{12}+\mathcal{J}^{\rm{int}}_2G^*_{21}\right)dx\nonumber\\
                  &+&2{\Re}\int\left( \frac{\mathcal{J}^{\rm{int}}_1\mathcal{J}_1^*}{P_1\alpha_1T_1}+\frac{\mathcal{J}^{\rm{int}}_2\mathcal{J}_2^*}{P_2\alpha_2T_2}+
                          \frac{\abs{\mathcal{J}_1}^2}{P_1\alpha_1T_1}+\frac{\abs{\mathcal{J}_2}^2}{P_2\alpha_2T_2}\right)dx,\nonumber\\
\eeA
where we have discarded the constant term $\mathcal{I}_3$. We can see here that those integrals have respectively the same structure as the entropy 
flow and entropy production derived in 
Refs.\cite{tome10,borlenghi17,borlenghi18}. In off-equilibrium steady states the quantity 
$\dot{\mathcal{M}}$ vanishes, thus the information flow is equal to minus the information production, up to the constant term. 
Therefore, in analogy with Refs.\cite{tome10,borlenghi17,borlenghi18}, we identify the information flow, i.e. the rate at which information is transferred between the two systems, with the first integrals of Eq.(\ref{eq:mutualinfo2}). The two terms of this first integral account respectively for the transfer of information from system 1 to system 2 and backwards, and we denote them by 
$\mathcal{T}_{12}$ and $\mathcal{T}_{21}$ correspondingly. Thus, the total information flow $\mathcal{T}=\mathcal{T}_{12}+\mathcal{T}_{21}$ is symmetrical upon exchange of the indexes 1 and 2 as it should, while the partial flows $\mathcal{T}_{ij}$, $i,j=1,2$ are the relevant observables that account for the directional propagation of information.

In order to obtain $\mathcal{T}_{ij}$, we insert the expressions of probability currents Eq.(\ref{eq:probcurrents2}), integrate by parts and substitute the integrals over $P_{12}$ in Eq.(\ref{eq:mutualinfo2}) with ensemble averages. a straightforward calculation, similar to the one performed in Refs.\cite{tome10,borlenghi17,borlenghi18} gives the following:
\be
\label{eq:infoflow}
\mathcal{T}_{ij}=\frac{\langle |G_{ij}|^2\rangle}{\alpha_iT_i}+2{\Re} \frac{\langle F_iG_{ij}^*\rangle}{\alpha_iT_i}+2{\Re}\langle \der_iG_{ij}\rangle.
\ee
Here one can see that the first term depends only on the square modulus of the coupling between the equations, while the second term contains the product of local forces and couplings. The last term comes from the products between the quantity $\der_iP_{12}$ and $G_{ij}$. It describes the direct effect of the bath and is obtained by applying the Stratonovich prescription for ensemble averaging, see Refs.\cite{tome10,borlenghi17} for a thorough discussion. 

At this point we can extend the previous calculations to the multivariate case. In particular, we consider a system of coupled stochastic differential equations
\be
\label{eq:multieq}
\dot{\psi}_i=F_i+G_i(X)+\sqrt{\alpha_iT_i}\xi_i,
\ee
for $i=1,...N$. As in the previous case, $F_i$ is the local force, while $G_i(X)$ models the coupling between the equations.
Here we denote $\Psi=\{\psi_1,...,\psi_n\}$ the ensemble of all the dynamical variables, $X\subset\Psi$ an arbitrary subset of $\Psi$ and $|X|$ is the number of element in $X$.
This permits to encode concisely all types of coupling between the equations, and not just a binary coupling. For example, a term like $G_1(X)$, with $X=\{\psi_2,\psi_3\}$ describe the coupling of equation 1 with equations 2 and 3. The FP equation associated Eq.(\ref{eq:multieq}) reads

\beA
\label{eq:multifp}
\dot{P}_X &=& \sum_{i=1}^{|X|}\{-\der_i[(F_i+G_i(X))P_X]-\der_i[(F_i^*+G_i^*(X))P_X]\nonumber\\
                &+&2\alpha_iT_i\der_i\der_i^*P_X  \},
\eeA
where $P_X$ refers to the probability distribution restricted to the subset $X$ and we denote by $P_\Psi$ the probability distribution for all the coupled equations.
Using a similar notation, the previous FP equation reads, in terms of probability currents, $\dot{P}_x=-2{\Re}\sum_{i=1}^{|X|}\der_i\mathcal{J}_i^X$, 
with $\mathcal{J}^X_i =F_i+G_i(X)P_X-D_i\der_i^*P_X$, an expression which generalises in a straightforward way Eqs.(\ref{eq:probcurrents2}).

To generalise to the multivariate case, we adopt the definition of mutual information for the ensemble $\Psi=\left\{\psi_1,\psi_2,...\psi_N\right\}$ developed by 
Fano \cite{fano61}  and reformulated by Han \cite{han80}. A synthetic but comprehensive review on the subject can be found in Ref.\cite{sirinvasa05}:

\be
\label{eq:multiinfo}
\mathcal{M}_N\equiv\mathcal{M}(\psi_1,...,\psi_N)=\sum_{i=1}^N(-1)^{i-1}\sum_{X, |X|=i}\mathcal{S}(X),
\ee
where $\mathcal{S}(X)=\int P_X\ln P_Xdx$ is the information entropy of subset $X$ and the phase space volume element here reads 
$dx=\left(\frac{i}{2}\right)^{|X|}\prod_{i=1}^{|X|}d\psi_i\wedge d\psi_i^*$. Expanding out Eq.(\ref{eq:multiinfo}) gives 
$\mathcal{M}_N=\mathcal{S}(\psi_1)+\mathcal{S}(\psi_2)+...+\mathcal{S}(\psi_N)-\mathcal{S}(\psi_1,\psi_2)-\mathcal{S}(\psi_1,\psi_3)  -...+...(-1)^{N-1}\mathcal{S}(\psi_1,...,\psi_N)$,
which contains all possible combinations of the $\psi$s. 
We now calculate the time derivative of each term of the total information entropy. Precisely as Eqs.(\ref{eq:mutualinfo1}) and (\ref{eq:mutualinfo2}), 
each member of the expansion contains the three terms: information flow, information production and a constant. Thus, the information flow reads:
\be
\mathcal{T}=\sum_{k=1}^N(-1)^{k-1}\sum_{X,|X|=k}\sum_{i=1}^{|X|}2{\Re}\int\mathcal{J}_i^XG_i^*(X)dx
\ee
where the partial flow that accounts for the transfer between oscillator $i$ and the oscillators of subset $X$ read simply $\mathcal{T}_{iX}=2{\Re}\int\mathcal{J}_i^XG_i^*(X)dx$.
Upon substituting the explicit expressions for the currents $\mathcal{J}_i^X$ and ensemble-averaging leads to the following expression:
\be
\label{eq:multiflow}
\mathcal{T}_{iX}=\frac{\langle |G_i(X)|^2\rangle_X}{\alpha_iT_i}+2{\Re} \frac{\langle F_iG_i^*(X)\rangle_X}{\alpha_iT_i}+2{\Re}\langle \der_iG_i(X)\rangle_X\
\ee
where the average of a function $f$ on the subset $X$ is defined as $\average{f}_X=\int f P_Xdx$.
One can see that this is a straightforward extension of the two systems case described in Eq.(\ref{eq:infoflow}).

As a simple example of information transfer, we consider here the dynamics of two coupled nonlinear oscillators, a model which implements the simplest
possible realisation of the discrete nonlinear Schr\"odinger equation (DNLS) \cite{iubini13} and has been applied to a variety
of physical systems, including coupled spin transfer nano oscillators \cite{slavin09} and the spin-Seebeck diode \cite{borlenghi14a,borlenghi14b}:

\beA
\label{eq:dnls}
\dot{\psi}_1 &=& (i-\alpha_1)(\omega_1\psi_1+A\psi_2)+\mu_1\psi_1+\sqrt{\alpha_1T_1}\xi_1,\nonumber\\
\dot{\psi}_2 &=& (i-\alpha_2)(\omega_2\psi_2+A\psi_1)+\mu_2\psi_2+\sqrt{\alpha_2T_2}\xi_2,
\eeA
with $\psi_i(t)=\sqrt{p_1(t)}e^{\phi_i(t)}$ , $i=1,2$. The nonlinear frequencies and damping
are respectively $\omega_i(p_i)=\omega_i^0(1+qp_i)$ and $\alpha_i(p_i)=\alpha_i^0\omega_i$. Here $q$ is the nonlinearity coefficient and $(\omega_i^0,\alpha_i^0,T_i)$ are respectively the linear frequency, damping and temperature of the bath. Note that in STNOs the frequency is proportional to the external magnetic field applied on the sample, and can be easily be controlled. For brevity in the following we will not write explicitly the dependence on the powers $p_i$. The chemical potential $\mu_i$ is a control parameter that acts as a gain that opposes to the damping. In STNOs it corresponds to the intensity of spin transfer torque, proportional to the injected electrical current, that excites the dynamics of the magnetisation.
Note that we consider a dissipative coupling $(i-\alpha_i)A$, with $A$ a real number modelling the strength of the coupling. This ensures that the system reaches thermal equilibrium when temperatures and chemical potentials are the same, as it has been discussed in Refs.\cite{iubini13,borlenghi15}.

At this point we insert Eqs.(\ref{eq:dnls}) into the definition of information flow Eq.(\ref{eq:infoflow}). By noting that the local forces read $F_i=(i-\alpha_i)\omega_i\psi_i+\mu_i\psi_i$ and the coupling $G_{ij}=(i-\alpha_i)A\psi_j$ for $i,j=1,2$, we obtain:
 
\beA
\label{eq:infoosc1}
\mathcal{T}_{12} &=& \frac{1+\alpha_1^2}{\alpha_1T_1}A^2\average{p_2}\nonumber\\
                            &+&\frac{2}{\alpha_1T_1}{\Re}\average{[i\omega_1-(\alpha_1-\mu_1)](i-\alpha_1)A\psi_1\psi_2^*},\nonumber\\
\eeA
We remark that the last term of Eqs.(\ref{eq:infoflow})  and (\ref{eq:multiflow}), proportional to $\der_iG_{ij}$, is zero in the case considered here of linearly coupled oscillators.
The first and second term of Eq.(\ref{eq:infoosc1}) are the incoherent and coherent component, to which we shall refer respectively as $\mathcal{T}_{12}^I$ and $\mathcal{T}_{12}^C$. As discussed before, the incoherent component accounts for the transfer of information due to the powers, while the coherent component is proportional to the phase difference 
$\Delta\phi=\phi_1-\phi_2$ between the oscillators and therefore depends on their phase synchronisation. This can be seen by writing the coherent component in terms of the phases and powers
 as $\mathcal{T}_{12}^C=\average{A(p_1,p_2)\sin\Delta\phi-B(p_1,p_2)\cos\Delta\phi}$, with $A(p_1,p_2)=\left[\frac{2(1+\alpha^2)\omega_1}{\alpha_1T_1}+2\mu_1\right]\frac{A\sqrt{p_1p_2}}{T_1}$ and $B(p_1,p_2)=\frac{2\mu_1}{\alpha_1T_1}A\sqrt{p_1p_2}$. We note however that the incoherent component is not completely independent on the phases, since the equations for phase and powers are coupled. It means simply that it can be nonzero even in the absence of phase synchronisation, when the powers are different.

The relevant observables of the coupled oscillators out of equilibrium are the differences of incoherent and coherent information flows $\Phi^{I/C}={\mathcal{T}_{12}^{I/C}}-{\mathcal{T}_{21}^{I/C}}$, which account for the net transfer of information between the oscillators. Other important observables are the particle and energy currents, respectively $j_p=2{\Im}\average{A\psi_1\psi_2^*}$ and $j_E=2{\Re}\average{A\psi_1\dot{\psi_2}^*}$, which accounts for the transfer of the powers $p_i$ and energy between the oscillators. The derivation of those current is done by calculating the time derivative of the powers $p_i$ and of the Hamiltonian of the system, and has been performed in great details in Refs. \cite{iubini13,borlenghi14a,borlenghi14b}, to which we refer for a thorough discussion. Under the condition considered here, $j_p$ and $j_E$ have the same profile and similar behaviour up to a scaling factor proportional to the frequency of the oscillators \cite{borlenghi14b}. Thus, for brevity here we report only the analysis of $j_p$. We remark that in spin systems the latter corresponds to the spin wave current that describes the net transport of the magnetisation between neighbouring macrospins \cite{borlenghi14b}  and as with 
$\Phi^C$, it is a coherent quantity proportional to the sine of the phase difference between the oscillators \cite{borlenghi14a,borlenghi14b}.

To better illustrate the information transfer in different off-equilibrium situations, we turn now tu numerical simulations. Eqs.(\ref{eq:dnls}) where integrated by means of a fourth order Runge-Kutta method, with an integration time step $dt=0.05$, coupling $A=0.01$, linear damping coefficient $\alpha_1^0=\alpha_2^0=0.01$, linear frequencies $\omega_1^0=1$ and 
$\omega_2^0=2$ and nonlinearity coefficient $q=2$. 

At first, we study the effect of chemical potential differences by considering the same temperatures $T_1=T_2=0.1$ model units. Starting from the condition of thermal equilibrium with 
$\mu_1=\mu_2=0.01$, we increase separately $\mu_1$ and $\mu_2$ and calculate the observables as a function of $\Delta\mu=\mu_1-\mu_2$. The equations where evolved for $5\times 10^5$ time steps, averaging the observables over the last $3.5\times 10^5$ time steps, where the system is in a steady state.
We remark that the setup described here behaves as a spin-Seebeck diode \cite{borlenghi14a,borlenghi14b}: since the frequencies are nonlinear and depend on the powers, changing 
$\mu_i$, $i=1,2$ allows one to control the phase synchronisation between the oscillators, moving from a desynchronised regime with where $\Phi^C$ and $j_p$ are close to zero  to a synchronised regime where those two observable strongly increase with $\Delta\mu$. In fact, in the desynchronised regime  the phases $\phi_1$ and $\phi_2$ of the two oscillators are not locked and evolve independently in time, so that the quantity $\Delta\phi=\phi_2-\phi_1$ is not constant and the terms containing $\sin\Delta\phi$ and $\cos\Delta\phi$ oscillate around zero and vanish in average. This means that the current moves back and forth between the oscillators, and there is no net transport. On the other hand, in the phase-locked regime $\Delta\phi$ approaches a constant value and the current is not zero in average.

\begin{figure}
\begin{center}
\includegraphics[width=8cm]{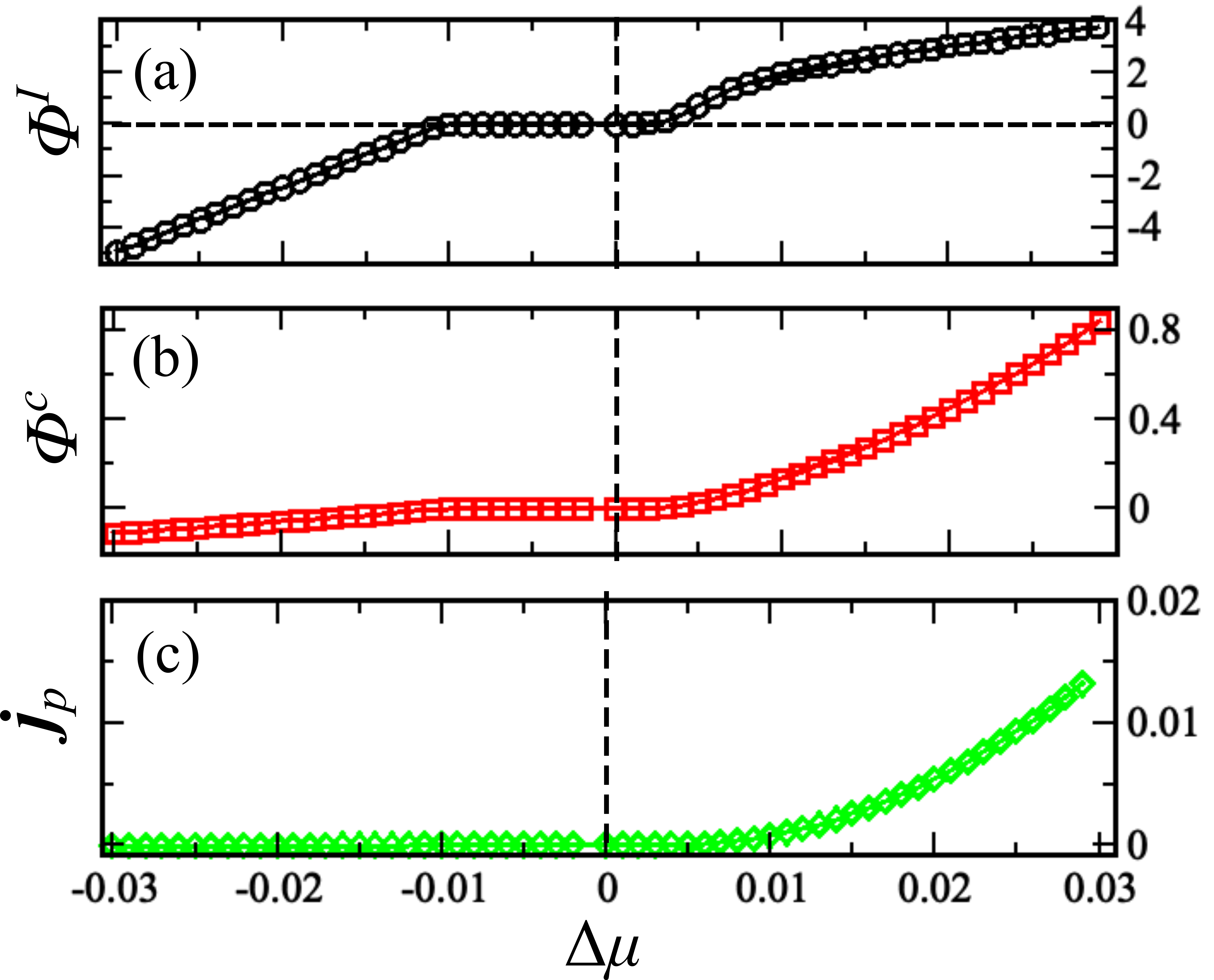}
\end{center}
\caption{Panels a) and b) show the incoherent and coherent component of the information transfer. While $\Phi^I$ changes sign and increases in magnitude with $\Delta\mu$, $\Phi^C$ increases with $\Delta\mu > 0$ and remains close to zero with $\Delta\mu<0$, showing a strong rectification effect. c) The particle current $j_p$ also displays a rectification effect, with its profile similar to that of $\Phi^C$.}
\label{fig:figure1}
\end{figure}

This behaviour can be seen in Fig.\ref{fig:figure1}, where the panels a) and b) show respectively $\Phi^I$ and $\Phi^C$. One can see that the incoherent component, with depends only on the difference in amplitudes of the two oscillators, always increases in amplitude with $\Delta\mu$. On the other hand, the coherent component displays a strong rectification effect, being close to zero when $\Delta\mu<0$ and increasing strongly when $\Delta\mu>0$. Panel c) shows the behaviour of the particle current $j_p$, which being a quantity that depends on the phase synchronisation also displays a strong rectification effect. 

\begin{figure}
\begin{center}
\includegraphics[width=8cm]{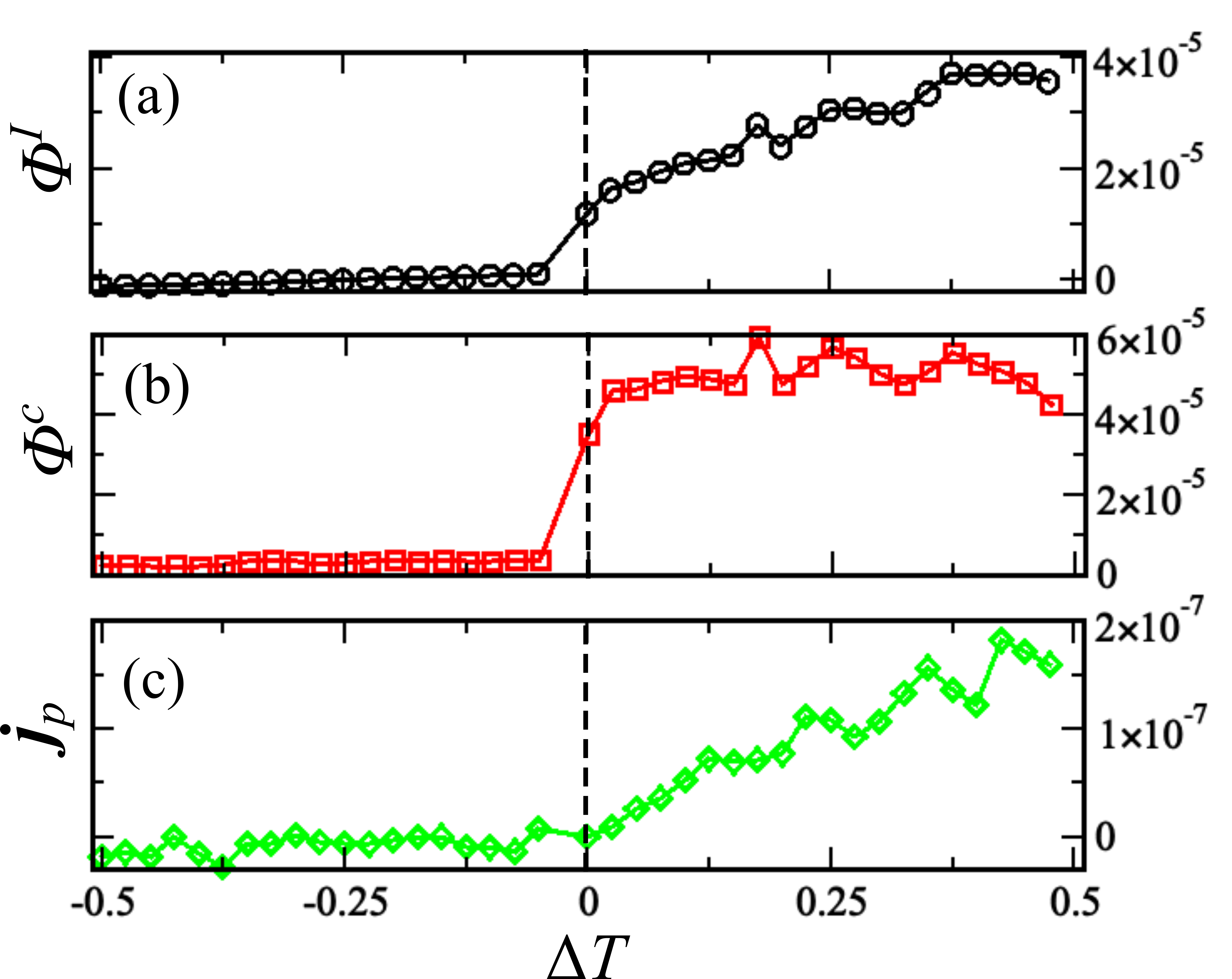}
\end{center}
\caption{Panels a) and b) show the incoherent and coherent component of the information transfer, which both are close to zero when
 $\Delta T < 0$ and increase with $\Delta T>0$. Together with the the particle current $j_p$ shown in panel c), they display a strong rectification effect}
\label{fig:figure2}
\end{figure}

Next, we consider the effect of temperature difference $\Delta T=T_2-T_1$ on the information and particle flows. To this end, we integrate the two oscillators Eqs.(\ref{eq:dnls}) by considering the same parameters as before, except for $\mu_1=\mu_2=0$ and the two temperatures that are different. In particular, we consider the system at thermal equilibrium with $T_1=T_2=1$ model units, and we increase separately $T_1$ and $T_2$, averaging the flows over  $3\times10^7$ time steps. The observables are reported in Fig.\ref{fig:figure2},
where one can see that all the currents display a strong rectification effect.

Thus, our simulation show two remarkable aspect of information transfer out of thermal equilibrium: at first, the information flow can be rectified. Then, under certain conditions it is possible to transfer incoherent information without transferring energy. In this respect, the behaviour of the information flow is quite different and more complex than the behaviour of the other currents. In the Spin-Seebeck diode, $\Delta\mu$ and $\Delta T$  have similar effects in rectifying $j_p$ and $j_E$, however this is not the case for $\Phi^I$.

In summary, we have derived a simple and general analytical expression to calculate the flow of information between an arbitrary number of coupled physical systems out of thermal equilibrium. At variance with the transfer entropy formalism, our formulation shows the physical origin and che characteristic of the information transfer, which depends on local forces and coupling between the equations, and in coupled oscillators contains both coherent and incoherent components. 

The formulation presented here is very general and has applications in several areas of Physics and technology. Oscillator networks models permit study the flow and storage of information in a variety of physical systems, such as nano-phononics \cite{lepri03,balandin12}, STNOs and spin-Josephson devices \cite{slavin09,borlenghi15b}. Other systems such as photonics waveguides, photosynthetic reactions, Bose-Einstein condensate \cite{kevrekidis01}, chaotic and chimera states in coupled oscillators \cite{abrams04} and electrical power grids can be investigated.  More exotic situations such as discrete breathers and negative temperature states  \cite{iubini13b}, dephasing-assisted spin transport \cite{iubini20} and anomalous heat transport in oscillator chains \cite{lepri97} involve information production and sharing and can be succesfully studied with our formalism.
Recently STNOs and coupled oscillators have been used to perform neuromorphic computing \cite{torrejon17,borlenghi18}, and our formalism allows to establish the amount of information that can be processed in such devices.
Finally, we remark that oscillator networks described by the DNLS present $U(1)$ gauge invariance, and our formalism allows to establish a connection between information transport and lattice gauge theories \cite{borlenghi16}. Several of these arguments will be studied in forthcoming papers. We wish to thank Dr. S. Iubini for useful comments and for reviewing the manuscript. 


\end{document}